\documentclass[aps,prd, preprintnumbers,showpacs,showkeys,nofootinbib,superscriptaddress]{revtex4}
\usepackage{epsfig,graphicx}
\usepackage{amssymb,amsmath,amsfonts,amsthm}
\usepackage{multirow}
\usepackage[dvipsnames]{xcolor}
\usepackage[bookmarks=false]{hyperref}
\hypersetup{pdfstartview={XYZ null null 1.25}}
\usepackage[normalem]{ulem}

\graphicspath{{./plots/}}                 
\newcommand{\psibar}{\ensuremath{\bar\psi}}
\newcommand{\chibar}{\ensuremath{\bar\chi}}
\newcommand{\pbp}{\ensuremath{\psibar\psi}}
\newcommand{\pbpvac}{\ensuremath{\langle \psibar\psi \rangle}}

\newcommand{\Nt}{\ensuremath{N_\tau}}
\newcommand{\Ns}{\ensuremath{N_\sigma}}
\newcommand{\vev}[1]{\ensuremath{\left<#1\right>}}

\newcommand{\Real}{\ensuremath{\mathrm{Re}}}

\newcommand{\Trace}{\ensuremath{\mathrm{Tr}}}
\newcommand{\Tr}{\ensuremath{\mathrm{Tr}}}

\newcommand{\Eq}[1]{Eq.\,(\ref{#1})}

\newcommand{\ie}{i.~e.\ }

\def\lsi{\raise0.3ex\hbox{$<$\kern-0.75em\raise-1.1ex\hbox{$\sim$}}}
\def\gsi{\raise0.3ex\hbox{$>$\kern-0.75em\raise-1.1ex\hbox{$\sim$}}}

\newlength{\graphicswidth}
\begin{document}
\preprint{HU-EP-14/64,~SFB/CPP-14-103}

\title{Chiral observables and topology in hot QCD with two families of quarks}

\author{Florian Burger}
\affiliation{Humboldt-Universit\"at zu Berlin, Institut f\"ur Physik, 
 12489 Berlin, Germany}

\author{Ernst-Michael Ilgenfritz}
\affiliation{Joint Institute for Nuclear Research, BLTP, 
141980 Dubna, Russia} 

\author{Maria Paola Lombardo{\footnote{Address after October, 1st:
INFN sezione di Firenze, Via G. Sansone 1; I-59100 Sesto F.no, Italy}}}
\affiliation{Laboratori Nazionali di Frascati, INFN, 100044 Frascati, Roma, Italy}

\author{Anton Trunin} 
\affiliation{Samara National Research University, 443086 Samara, Russia}

\collaboration{tmfT Collaboration}
\noaffiliation

\date{August 19, 2018}


\begin{abstract}
We present results on QCD with four dynamical
flavors in the temperature range $150$ MeV $\lesssim T \lesssim 500$ MeV.
We have performed lattice simulations with
Wilson fermions at maximal twist and measured 
Polyakov loop, chiral condensate and disconnected susceptibility,
on lattices with spacings as fine as 0.065~fm. 
For most observables spacing effects are below statistical errors, 
which enables us to identify lattice results with continuum estimates. 
Our estimate of the pseudocritical temperature compares favorably with
continuum results from staggered and domain wall fermions, confirming that a dynamical charm does not 
contribute in the transition region.
From the high temperature behaviour of 
the disconnected chiral susceptibility we infer the topological
susceptibility, which encodes relevant 
properties of the QCD axion, a plausible Dark Matter candidate.
The topological susceptibility thus measured exhibits a power-law decay
for $T/T_c \gtrsim 2$, with an exponent close to the one predicted
by the Dilute Instanton Gas Approximation (DIGA).
Close to $T_c$ the temperature dependent effective exponent seems to approach 
the DIGA result from above, a behaviour which would support 
recent analytic calculations based on an Instantons-dyons model.
These results constrain the mass of a hypothetic QCD post-inflationary 
axion, once an assumption concerning the relative contribution of axions 
to Dark Matter is made.
\end{abstract}

\maketitle

\section{Introduction} \label{sec:intro}

The properties of strong interactions at high temperatures are under 
active scrutiny both theoretically and experimentally. From an 
experimental viewpoint, collisions of ultrarelativistic ions at the 
LHC create matter whose temperature has been estimated to reach 
500 MeV and more. Theoretically, the behaviour of the theory at high 
temperature offers important insights into the mechanisms of chiral 
symmetry, confinement and the degrees of freedom active in the different
phases. 
For instance, both experiments and lattice results indicate that the 
system remains very strongly coupled above the critical temperature 
$T_c$, and that the 
perturbative regime, and eventually a free field behavior, will only 
set in at much higher temperatures, see e.g. Ref.~\cite{Ding:2016qdj}.

As a consequence of this, fundamental degrees of freedom manifest 
their influence in a sequential way when temperature increases: 
particles of the light spectrum dissolve first, heavy bound states 
later. Among the quarks themselves, the light ones, which receive 
most of their masses at the quark-gluon transition, have a significant
influence on the transition itself: the dependence of the value of 
the pseudocritical  temperature(s) as well as of the nature of 
the transition on the masses of light current quark masses, 
as well as the role of the strange quark mass around $T_c$ have been 
studied at a great length, and the current status of our
understanding of the sensitivity of the
phase diagram to the light degrees of freedom is for instance 
summarized in the so-called Columbia plot~\cite{Schmidt,deForcrand,Cuteri}.
Closely related with the issue of nature of the phase transition is
the (effective) restoration of the $U_A(1)$ symmetry~\cite{Sharma:2018syt,Cossu}.

By contrast, we know that the charm 
quark mass does not carry any influence on the transition: if we were to draw   
a multidimensional Columbia plot including the charm mass $m_c$ the plot  
will just 
look the same for values of the charm mass ranging from infinity 
down to the experimental value. This can be anticipated just by considering that
the charm does not acquire its mass across the chiral phase transition,
and our numerical results will confirm this.  
However, as the temperature increases the role of the charm becomes 
more important, and HTL calculations predict a temperature threshold for 
sizeable charm effects at about 300 MeV~\cite{Laine:2006cp}.

This suggests that precise calculations of QCD thermodynamics in the
LHC working region, where temperature as high as 500 MeV may be reached,
require a dynamical charm. However, in comparison with
studies with $2+1$ flavors, lattice results with two dynamical families
are much less developed~-- lattice discretization effects are more severe due
to the charm large mass. Only the MILC collaboration~\cite{Bazavov:2013pra}
and the Wuppertal--Budapest collaboration~\cite{Borsanyi} 
have published results for the Equation of State with
$N_f = 2+1+1$ flavors based on the staggered action. Our own study
of the Equation of State is underway
and only preliminary and partial results have appeared~\cite{Burger:2015xda},
which will not be presented here. 
 
In this paper we study the chiral and topological properties of QCD for temperatures 
ranging from about 150 MeV till about 500 MeV, well above the deconfinement 
transition, and close to the charm threshold. We use four flavors 
of maximally twisted mass Wilson fermions in the isospin limit, 
\ie with degenerate up and down quark masses, and physical strange 
and charm masses, relying on the zero temperature results of the 
ETM collaboration for scale setting and tuning of the algorithm 
to simulate at maximal twist~\cite{tmfT-Nf=2+1+1-first}.

The purpose of this study is twofold: firstly, we present the first
results around the pseudocritical temperature obtained with twisted 
mass Wilson fermions. We successfully cross check our results with 
earlier ones obtained with $N_f=2+1$ staggered and domain wall
fermions~\cite{BW2010,hotqcd2012,hotqcd2014}, 
thus  confirming that a dynamical charm has no influence around
the transition. Secondly, we will use the results for the disconnected 
chiral susceptibility to calculate the topological susceptibility 
for temperatures as high as 500 MeV, apparently reaching the onset 
of the Dilute Instanton Gas Approximation. Since this second aspect 
is at the moment under active investigation by several groups, we 
conclude this introduction with a brief review of the current status 
of topology in hot QCD.

QCD topology is an eminently non perturbative problem which has been 
addressed on the lattice since long. It is well known that topological 
studies are hampered by technical difficulties on a discrete 
lattice~\cite{Muller-Preussker,bonati-topreview}. However, recent methodological 
progress, together with more adequate computer resources, have to some extent 
reopened the field, leading to the first results on topology at high 
temperature with dynamical
fermions~\cite{Trunin,Bonati,Borsanyi,Petreczky,Taniguchi}.
Although these studies exhibit some common features, which we will review 
in the following together with our own results, a quantitative agreement 
has not been reached yet. Particularly significant~-- and still under debate~--
is the onset of the Dilute Instanton Gas behaviour: once this is reached, the
results could be safely extrapolated to temperatures $T=\mathcal O(1)$~GeV of cosmological relevance. 

In the calculation of the topological susceptibility  presented here
we  follow an early proposal of Kogut, Lagae, and 
Sinclair~\cite{KLS} which has also been investigated
by other groups~\cite{Bazavov,Petreczky,Buchoff}. In a nutshell, we will
use well known identities in the fermionic sector based on the Atiyah--Singer theorem~--
as we will review
in the following~-- to infer 
the topological susceptibility at high temperatures from the results
for the chiral susceptibility.

Let us finally mention that  Ref.~\cite{Wantz:2009it} proposed to
use lattice results as a quantitative input to axion cosmology. Berkowitz, Buchoff and
Rinaldi~\cite{Berkowitz} were the first ones to implement this suggestion in a paper which we
regard as seminal, as it has  inspired a large lattice activity focused on axion cosmology
(despite being based on the quenched model). Not surprisingly, given the
exploratory nature of these studies  and the already mentioned
complications of lattice topology,   
the results on axion cosmology 
have not reached an unanimous consensus--
even pure Yang Mills is still under investigation, see e.g. Refs.~\cite{Borsanyi:2015cka,Jahn:2018dke}.
This calls for further studies, thus providing one of the motivations of this work.

This paper is organised as follows. In the next Section we review the lattice
action and the setup of the simulations. Observables are introduced in Section III. 
The following two Sections contain our results:  Section IV is devoted to the
analysis of  the pseudocritical region, while in Section V we discuss topology
and its implication on the bounds on the post-inflationary QCD axion. 
We close with a brief discussion on the present status of $N_f=2+1+1$ thermodynamics
and topology, and future steps. 

Some of the results presented here have been presented in a preliminary form at
Conferences~\cite{Burger:2017xkz,Burger:2015xda}.

\section{The action and the simulation setup}

We performed simulations with four flavors of maximally twisted mass Wilson 
fermions in the isospin limit, \ie with degenerate, larger than physical,
up and down quark masses, and physical strange and charm quark masses.
In terms of the twisted fields 
$\chi_{l,h}  = \exp{(- i \pi \gamma_5 \tau^{3} / 4 )} \psi_{l,h}$
the light and heavy quark twisted mass actions have the following form:
\begin{equation}
S^l_f[U,\chi_l,\chibar_l] = \sum_{x,y} \chibar_l(x) \left
 [\delta_{x,y} -\kappa D_\mathrm{W}(x,y)[U] + 2 i \kappa a \mu \gamma_5 \delta_{x,y}  \tau^3 \right ]
 \chi_l(y) \;,
\label{tmaction}
\end{equation}
and similarly:
\begin{equation}
  S^h_f[U,\chi_h, \overline{\chi}_h] =\sum_{x,y} \overline{\chi}_h(x) \left[
\delta_{x,y} - \kappa D_W(x,y)[U] + 2 i \kappa a \mu_{\sigma} \gamma_5 \delta_{x,y}\
 \tau^1 + 2 \kappa a \mu_{\delta} \delta_{x,y}
\tau^3 \right] \chi_h(y) \;.
\label{eq_heavyaction}
\end{equation}
where $D_\mathrm{W}[U]$ is the usual Wilson operator, $a$ is the lattice spacing,
and $\chi_{l,h}$ are quark spinors in the twisted basis. The hopping parameter 
$\kappa$
is set to its coupling dependent critical value $\kappa_c(\beta)$ leading to the
so-called ''maximal twist`` of the action~\eqref{tmaction}--\eqref{eq_heavyaction} with the property of
automatic $\mathcal O(a)$ improvement for expectation values of any
operator~\cite{Frezzotti-Rossi,shindler}. The parameter $\mu_l$ describes the mass of the
degenerate light quark doublet, which is still unphysically large in our study:
the charged pion mass values $m_{\pi^\pm}$ considered at present are
210, 260, 370 and 470~MeV. The heavy twisted mass parameters $\mu_{\sigma}$ and
$\mu_{\delta}$ have been tuned in the unitary approach to reproduce approximately
the physical $K$ and $D$ meson mass values within the accuracy of 10\%, thus allowing
for a realistic treatment of $s$ and $c$ quarks.

For the gauge sector the Iwasaki action is used 
($c_0 = 3.648$ and $c_1 = -0.331$):
\begin{equation}
S_g[U] = \beta  \Big{(} c_0 \sum_{P} \lbrack 1 - \frac{1}{3} \Real \Trace \left( U_{P} \right ) \rbrack
 + c_1 \sum_{R} \lbrack  1 - \frac{1}{3} \Real \Trace  \left ( U_{R} \right ) \\
\rbrack \Big{)} \;.
\label{tlsym}
\end{equation}
The two sums extend over all possible plaquettes ($P$) and planar rectangles 
($R$), respectively.

Our finite temperature simulations have been performed for three 
values of $\beta =(1.90,1.95,2.10)$. Using the nucleon mass 
to fix the scale, this gives $a=0.0646~\mathrm{fm}$,
$a=0.0823~\mathrm{fm}$ and $a=0.0936~\mathrm{fm}$~\cite{Alexandrou}. 
For each lattice spacing we explored temperatures ranging from 150 MeV to 500 
MeV by varying the temporal size of the lattice $N_\tau$. 
So far we have generated finite temperature configurations for eight sets of 
parameters that correspond to four values of the charged pion mass of about 
470, 370, 260 and 210 MeV, for which two, three, two and one value(s) of the 
lattice spacing have been considered, respectively. 

For each lattice spacing 
the advantage of this setup (``fixed-scale approach''~\cite{umeda}) is that we are allowed 
to rely on the setup of $T=0$ simulations of ETMC, thus exploring 
 a wide set of temperatures. Different lattice spacings allow us to
study the approach to the continuum limit. In principle, we have
to deal with 
the mismatch of different temperatures obtained for different choices
of $\beta$.  In practice, the temperature scans are fine enough to
overcome this potential disadvantage, as we will see when presenting 
the results. The full set
of parameters, as well as the indicative statistics,
is reported in Table~\ref{tbl1}.

\begin{table}
{\small 
 \begin{tabular}{c|c|c|c|c|c}
tmft $T \ne 0$ & ETMC $T=0$   & $m_{\pi^{\pm}}$ [MeV] & $a$ [fm] & $N_\tau \times N_\sigma^3$                          & statistics\\
nomenclature   & nomenclature &  &  &  & \\
  \hline \hline
  D210 & D15.48         & 213(9)                   & 0.0646          & $\{4,6,8,10,12,14,16,18,20,24\} \times 48^3$        &  1k-7k \\
  \hline \hline  
  A260 & A30.32         & 261(11)                  & 0.0936          & $\{4,6,8,10,11,12,14\} \times 32^3$                  &  1k-5k \\
         &              &                          &                 & $\{16\} \times 40^3$ 				    &  3k    \\
       &                &                          &                 & $\{20\} \times 48^3$ 				    &  4k    \\
  \hline
  B260 & B25.32         & 256(12)                  & 0.0823          & $\{4,5,6,8,10,12,14,16,18\} \times 40^3$               &  1k-8k \\
  \hline \hline

  A370 & A60.24         & 364(15)                  & 0.0936          & $\{3,4,5,6,7,8,9,10,11,12\} \times 24^3$             &  2k-9k \\
       &                &                          &                 & $\{13,14\} \times 32^3$                              &  5k,27k \\
  \hline
  B370 & B55.32         & 372(17)                  & 0.0823          & $\{3,4,5,6,7,8,10,11,12,13,14,15,16\} \times 32^3$   &  2k-10k \\
B370.24&         &                & 		       & $\{4,6,8,10,11,12\} \times 24^3$                     &  3k-10k \\
  \hline
  D370 & D45.32         & 369(15)                  & 0.0646          & $\{5,6,7,8,9,10,12,14,16\} \times 32^3$              &  1k-12k \\
       &                &                          &                 & $\{18\} \times 40^3$                                 &     10k \\
       &                &                          &                 & $\{20\} \times 48^3$                                 &     10k \\       
  \hline \hline
  A470 & A100.24s       & 466(19)                  & 0.0936          & $\{4,5,6,7,8,9,10,11,12\} \times 24^3$               &  3k-8k \\
       &                &                          &                 & $\{14\} \times 32^3$                                 &  8k \\
  \hline
  B470 &B85.24          & 465(21)                  & 0.0823          & $\{4,5,6,7,8,9,10,11,12\} \times 24^3$               &  2k-4k \\
       &                &                          &                 & $\{13,14\} \times 32^3$                              &  2.5k,7k \\  
  \hline  
 \end{tabular}
 }
\caption{$T=0$ base ensembles, charged pion mass, lattice spacings 
and finite temperature ensembles generated so far. }
 \label{tbl1}
\end{table}

\section{Observables}

We concentrate first on the study of the critical region: our primary
observables here are the quasi order parameter for
deconfinement and chiral symmetry breaking~-- the 
Polyakov loop and  the chiral condensate in the light sector. We also
consider the disconnected chiral susceptibility: as it is merely
the fluctuations of the order parameter, it has the same meaning of
ordinary susceptibilities in spin models, hence 
it carries the relevant information on the pseudocritical behaviour.
We will also discuss later how this observable acts as a proxy for the
topological susceptibility in the symmetric phase.

\subsection{Polyakov loop}

The Polyakov loop is defined as a Wilson loop of gauge fields winding once 
around the temporal (thermal) direction. The important quasi order parameter
is the real part of it, 
\begin{equation}
\Real\left(L\right) = \frac{1}{N_c} \frac{1}{\Ns^3} 
           \Real \Trace \sum_{\mathbf{x}}\prod_{x_4=0}^{\Nt-1} U_4\left(\mathbf{x},x_4\right)\;.
\label{eq:polloop}
\end{equation}
We may only consider the real part since when quarks are present the center 
symmetry is explicitly broken towards the real sector. In pure gauge theory 
this quantity is strictly the order parameter for the deconfinement transition,
and since the pure gauge limit is approached for large quark masses the Polyakov
loop acts as a quasi order parameter also with dynamical quarks. It may be renormalized
 as follows~\cite{Aoki}
\begin{equation}
  \vev{\Real(L)}_R = \vev{\Real(L)}  \exp{(V(r_0)/2T)}\,,
\end{equation}
where $V(r_0)$ denotes the static quark-antiquark potential at the distance 
of the Sommer scale $r=r_0$~\cite{Sommer}. The latter is to be 
determined at zero temperature. 
The static quark-antiquark potential $V(r)$ has been evaluated using 20 steps
of APE smearing with a smearing parameter $\alpha_{\mathrm{APE}} = 0.5$.
For $r_0$ we use the values determined by ETMC.
Subsequently $V(r)$ was interpolated using cubic splines and its value $V(r_0)$
has been extracted. 

\begin{figure}
\includegraphics[width=0.49\linewidth]{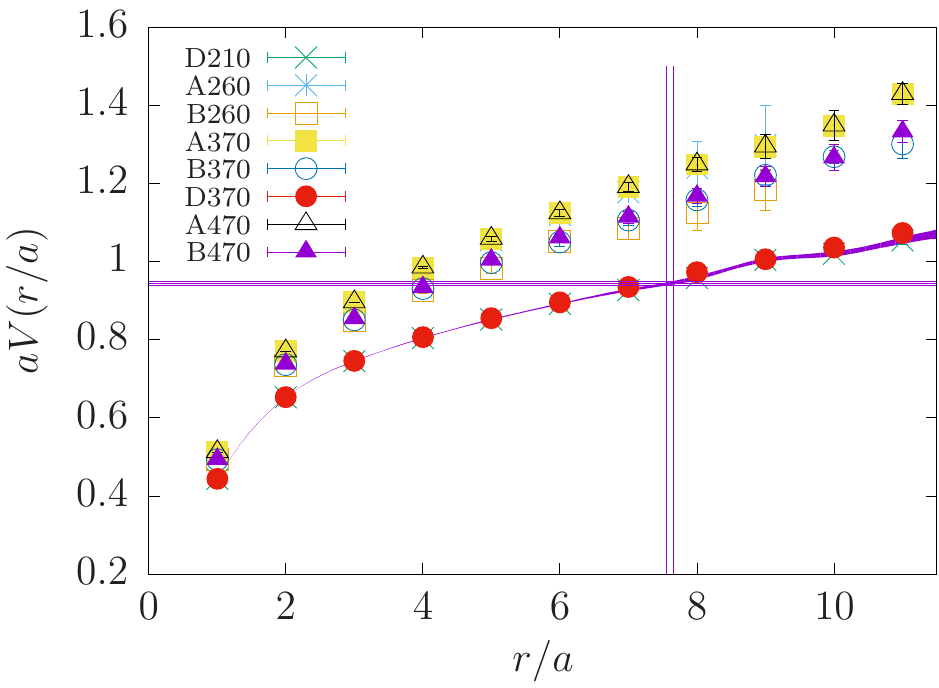}
\caption{The results for the potential at zero temperature used for the
renormalization of the Polyakov loop. The vertical band is drawn in correspondence of
the value of $r_0/a$ calculated by the ETMC collaboration for the D ensemble  and a 
pion mass of 210 MeV, hence the horizontal band indicates the corresponding renormalization
parameter $V_0$ which has been numerically determined as described in the main text. Analogous procedures have been followed for the other ensembles.}
\label{Fig:Potential}
\end{figure}

The results for the potential are summarized in Figure~\ref{Fig:Potential}
and the extraction of the value $V(r_0)$ 
needed for the renormalization of the Polyakov loop is esemplified for the 
results for a pion mass of 210 MeV on the D ensembles.  The others are  completely analogous.  
It is important to  note that this prescription~\cite{Aoki} introduces a further
dependence on temperature, hence in the crossover region we should not expect that the results
for the pseudocritical temperatures match those obtained using other temperature independent
renormalization prescriptions. Of course in the infinite mass limit, when the Polyakov loop
becomes an exact order parameter for deconfinement, these ambiguities disappear.

\subsection{Chiral condensate}

The  chiral condensate $\pbpvac$  in the light sector
serves as an (approximate) order parameter for the $SU(2) \times SU(2)$ symmetry,
which is explicitly broken by the quark mass: 
\begin{equation} 
 \pbpvac = \frac{T}{V} \frac {\partial \ln Z}{\partial m_q} = \frac{1}{N_\sigma^3N_\tau}\,\langle\, \Tr M_q^{-1}\rangle
\end{equation} 
The trace of the inverse 
is evaluated by means of the technique 
of noisy estimators using multiple quark matrix inversions acting on 24 
Gaussian noise vectors. At maximal twist  the leading
ultraviolet divergent part is proportional to 
${\mu}/{a^2}$ which in our early $N_f=2$ study 
was removed by subtracting the corresponding condensate 
at the same mass at $T=0$~\cite{Burger}. The multiplicative renormalization
factors are then canceled by dividing by the condensate at zero temperature
in the chiral limit:
\begin{equation}
R_{\pbpvac} = \frac{\pbpvac^{T,\mu}_l - \pbpvac^{T=0,\mu}_l + \pbpvac^{T=0,\mu=0}_l}{\pbpvac^{T=0,\mu=0}_l}\,.
  \label{eq_pbpratio}
\end{equation}

Having now the strange quark at our disposal, we are using another prescription based on a suitable subtraction (involving the light and strange masses) of 
the strange quark condensate. It eliminates the additive 
$\mu$-proportional divergence 
contained in the light condensate being estimated by means of the strange quark
condensate. This procedure has been proposed for $N_f=2+1$ flavors in the 
literature~\cite{Cheng}:
\begin{equation}
  \Delta_{l,s} = \frac{\pbpvac_l - \frac{\mu}{\mu_s}\pbpvac_s}{\pbpvac^{T=0}_l - \frac{\mu}{\mu_s}\pbpvac^{T=0}_s}\,,
   \label{eq_pbpsub}
\end{equation}
where $\pbpvac_s$ is the strange quark condensate calculated in the Osterwalder--Seiler setup~\cite{Osterwalder-Seiler,Frezzotti-Rossi} to avoid mixing in the heavy quark sector and where the strange mass $\mu_s$ has been determined as to reproduce the physical mass of $\bar s\gamma_\mu s$. The corresponding expression at $T=0$ serves as normalization factor.

\subsection{Chiral susceptibility}

The chiral susceptibility, defined as the mass derivative of the chiral condensate
reads:
\begin{equation}
\chi = \frac{V}{T} \frac{\partial}{\partial m_l} \vev \pbp_l \equiv \chi^{\rm disc}_{\pbp} + \chi^{\rm conn}_{\pbp}
\end{equation}
The disconnected chiral susceptibility, which carries the relevant information on
the critical behaviour, is the quadratic fluctuation of the chiral condensate:
\begin{equation}
\chi^{\rm disc}_{\pbp} = \frac{V}{T} \left(\vev{(\pbp)^2}_l - \vev{\pbp}_l^2 \right)\;.
\label{eq:sigma2}
\end{equation}
for which the traces arising from the path integral have been evaluated using 
the stochastic noise method and we made sure that no net connected piece enters 
the final result.
Ultraviolet divergences cancel in the difference,
and we are left only with a multiplicative renormalization.

\section{Thermal transition temperature(s)}

We have considered the renormalized Polyakov loop and two chiral observables 
in order to extract three different pseudo-critical temperatures. 
We will discuss the used fit strategies in what follows.

\subsection{From the Polyakov loop}

The deconfinement transition temperature $T_\mathrm{deconf}$ is read off 
the inflection point of a hyperbolic tangent function fit to the renormalized 
Polyakov loop data
\begin{equation}
\vev{\Real(L)}_R  =  A_P + B_P \tanh{\left ( C_P (T-T_\mathrm{deconf}) \right )} \;.
\label{eq:fit_P}
\end{equation}
The data becomes more and more noisy for larger $N_\tau$ which mostly reduces 
the data quality for the small mass ensembles.

We have restricted the fits to temperatures below $T=310 \mathrm{~MeV}$ for our 
central fits in order to avoid the region of visible discretisation 
artifacts at high temperatures. We have always included into the fits all 
available data at the low temperature end since data is very limited there. 

A second fit was performed including higher temperature data up to 
$T=380 \mathrm{~MeV}$ (or up to $T=400 \mathrm{~MeV}$ in the case of B260, 
respectively) in order to estimate a systematic error for $T_\mathrm{deconf}$ 
taking half of the deviation from the central fit results above.

\begin{figure}[htb] 
{\centering
\hfill
\includegraphics[width=0.33\linewidth]{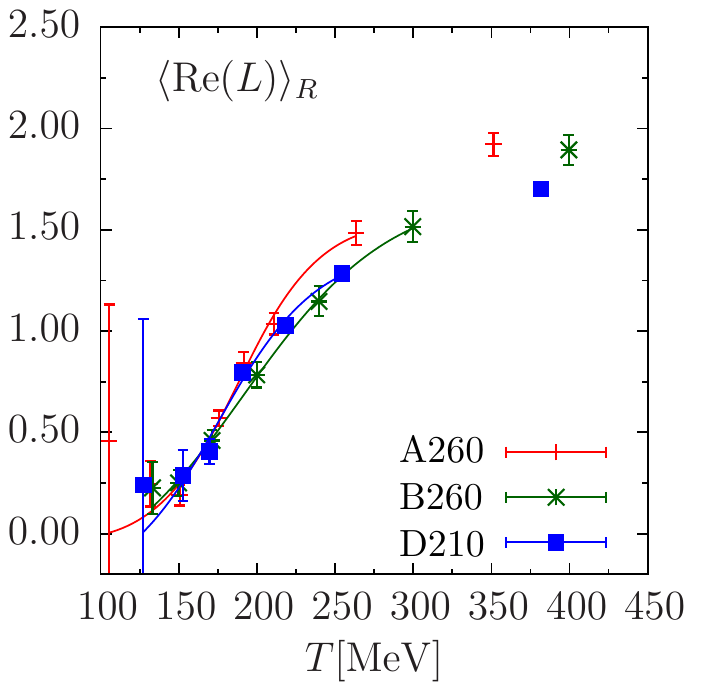}\hfill
\includegraphics[width=0.33\linewidth]{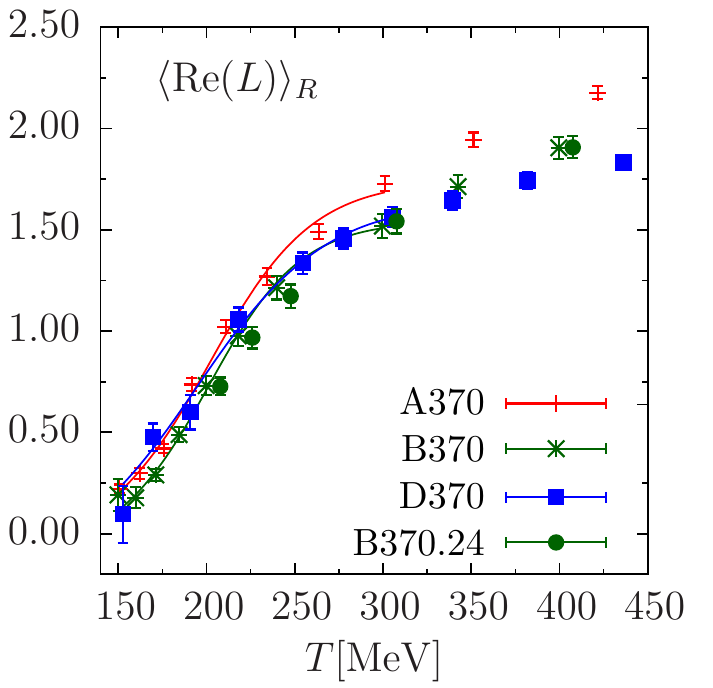}\hfill
\includegraphics[width=0.33\linewidth]{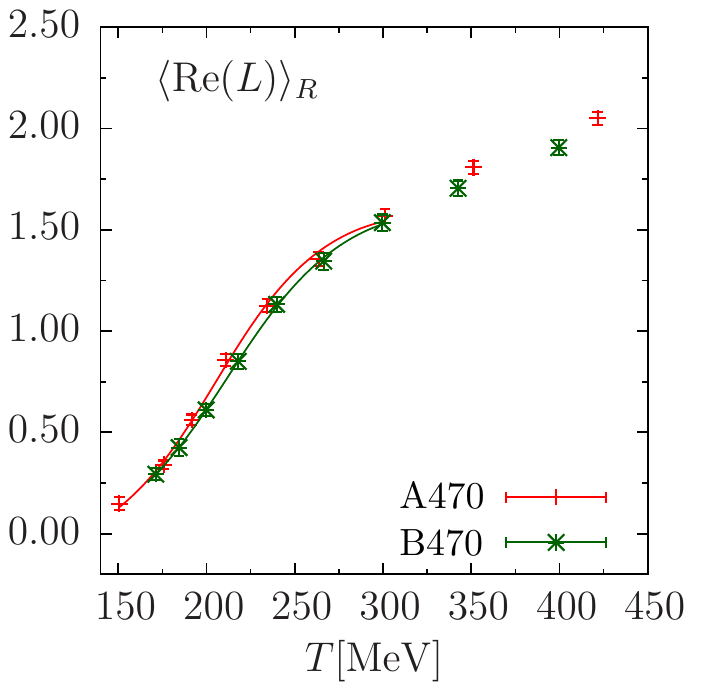}\hfill
}
\caption{The renormalized Polyakov loop. Left: for $m_\pi = 210 $ MeV (blue 
points) and for $m_\pi = 260 $ MeV.
Middle: for $m_\pi = 370 $ MeV. The data of the finite size test ensemble 
B370.24 is shown slightly shifted to ease reading the figure. 
Right: for $m_\pi =470 $ MeV.}
\label{fig_poly}
\end{figure}

In the middle panel of Figure \ref{fig_poly} we have added the Polyakov loop 
data of our finite size test $24^3$ ensemble B370.24. 
There is no difference in the data as compared to the data obtained in the 
larger volume $32^3$.

 A we have already discussed at the end of Section IIIA,  this renormalization prescription 
is temperature dependent hence it affects the estimate of the pseudocritical
temperature in the crossover region.  This is just one of the many examples
of a  pseudocritical temperature which is not unique.
As an alternative estimate of the pseudocritical temperature for deconfimenent we 
may define $T^{\cal{F}}_\mathrm{deconf}$ as the inflection point of the free energy 
for static quark (or, equivalently, from the peak of the entropy for 
static quark). Consider the free energy ${\cal{F}} = - T \log {\Real(L)}_R $  where the
renormalization constant is only additive and does not influence the inflection point position.
We have performed hyperbolic tangent fits in the interval [130:300] MeV and we have estimated
the systematic error on the inflection point by 
varying the fit interval and/or the fit function (polynomial fits). 
As it is well known, the free
energy is rather smooth (due to the logarithm), for the lightest mass the fits
were not successful and in the other cases  the identification of the inflection point has 
 large errors. All in all 
 $T^{\cal F}_\mathrm{deconf}$ is consistent with $T_\mathrm{deconf}$
within the errors. A high statistics study
with a  finer range of temperature will be necessary to reveal the expected shift between
 $T_\mathrm{deconf}$  and  $T^{\cal F}_\mathrm{deconf}$ 
which will anyway disappear when the Polyakov loop is an exact order parameter.

\subsection{From the renormalized chiral condensate}

Here the first estimate $T_\Delta$ for the crossover temperature $T_c$
is determined from the renormalized chiral condensate.
The data of $\Delta_{ls}$ follows an S-shape curve for all considered 
ensembles.  We therefore used as a fit ansatz the following sigmoid function:
\begin{equation}
\Delta_{ls}(T) = A_\Delta + B_\Delta \tanh \left ( -C_\Delta (T - T_\Delta) \right ) \;.
\label{eq:tanhfit}
\end{equation}
We always used all available data at low temperatures and used two upper limits 
for the fit ranges in temperature. The main fits were obtained with data with 
$T < 350$ MeV and another extended fit has been done with $T < 450$ MeV. 
Half the deviation of the latter from the main fit results was used to estimate 
the systematic error.

 Secondly, we have used a simple polynomial fit
\begin{equation}
\Delta_{ls}(T) = a_\Delta + b_\Delta T + c_\Delta T^2 + d_\Delta T^3 \;.
\label{eq:polyfit}
\end{equation}
we started from an exact fit using four points and we progressively enlarged the interval
till the quality of the fit deteriorates. The inflection point is then 
$T^p_\Delta = -c_\Delta/(3 d_\Delta)$ and one first estimate of the error is given by the 
maximum between the statistical errors of the individual fits, and the dispersion of the
results obtained by changing the interval as described. 
Taking into account that the polynomial fit merely serves as an interpolator in the given
interval,  a second more conservative estimate for the error 
uses the resolution of the interpolation - in practice, 
 half of the distance between the simulations point which are closest to $T_\Delta$.
We quote both errors in the table, where we also keep both results, from the sygmoid fit and
from the polynomial fit, as the dispersion between them offers a further estimate of the
errors. It turns out that within the larger errors associated with the polynomial fits
$T_\Delta$ and $T^p_\Delta$ agree.

\begin{figure}[htb] 
{\centering
\hfill
\includegraphics[width=0.33\linewidth]{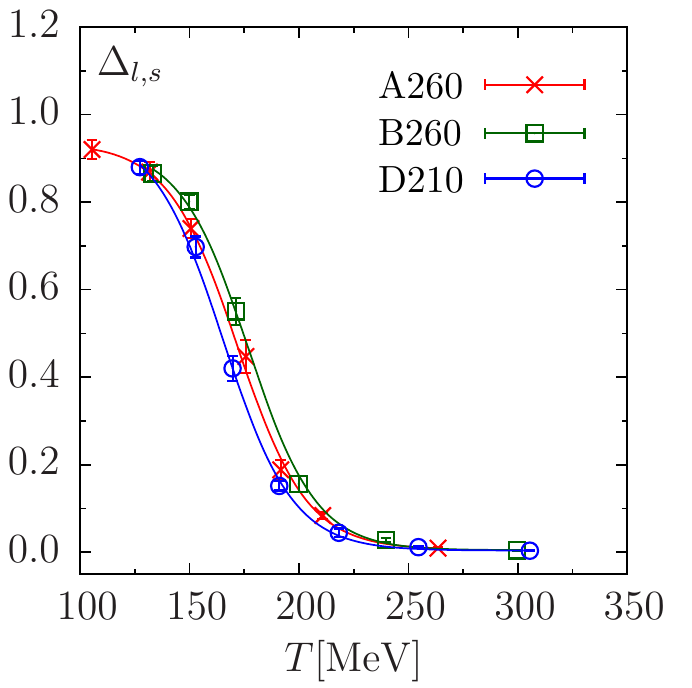}\hfill
\includegraphics[width=0.33\linewidth]{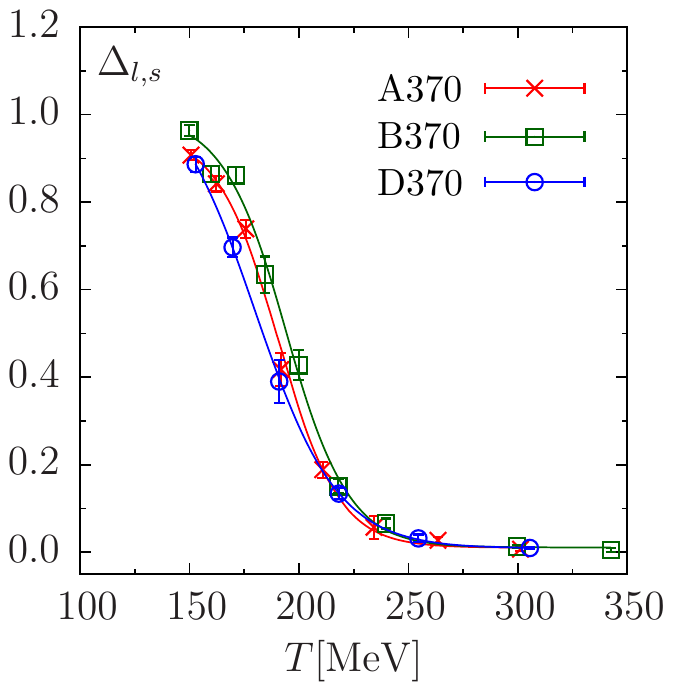}\hfill
\includegraphics[width=0.33\linewidth]{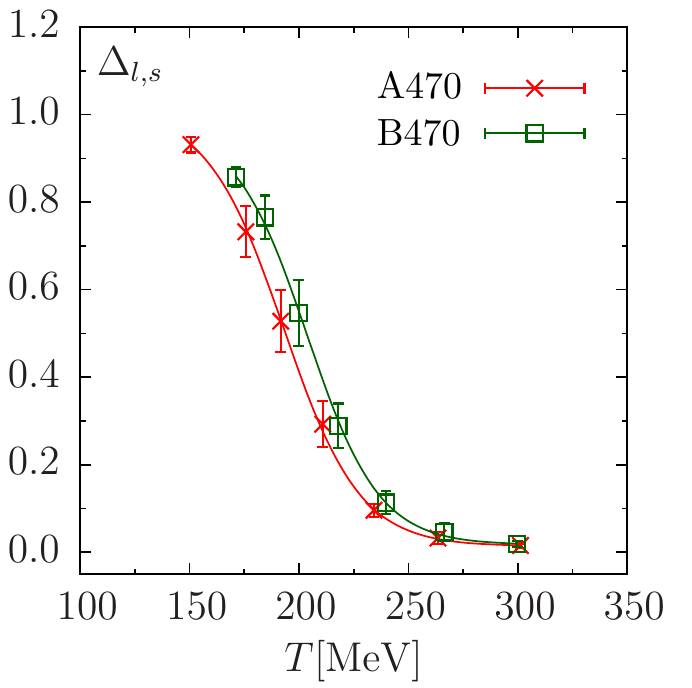}\hfill
}
\caption{The renormalized chiral condensate $\Delta_{ls}$. Left: for 
$m_\pi = 210 \mathrm{~MeV}$ (blue points) and for 
$m_\pi = 260 \mathrm{~MeV}$.  Middle: for 
$m_\pi = 370 \mathrm{~MeV}$.  Right:  for 
$m_\pi = 470 \mathrm{~MeV}$.}
\label{fig_pbpren}
\end{figure}

\subsection{From $T\chi^{\rm disc}_{\bar\psi\psi}$}

Here the second estimate for the crossover temperature 
$T_\chi$ is determined from the chiral susceptibility. 
We noted that by considering the  $T\chi^{\rm disc}_{\bar\psi\psi}$ rather
than $\chi^{\rm disc}_{\bar\psi\psi}$ 
the signal is sharper, and we used this in our estimates. The modest
shift in the pseudocritical temperature associated with this different
normalization is about $10$~MeV, negligible with respect to other errors.

For estimating the position of the peak in the (modified) disconnected chiral 
susceptibility \Eq{eq:sigma2} we fit an ansatz quadratic in the 
temperature to the data:
\begin{equation}
\chi^{\rm disc}_{\pbp}(T) = A_\chi + 
B_\chi \left ( T - T_\chi \right )^2 \;,
\label{eq:squarefit}
\end{equation}
a definition that does not assume a model of the $T$-dependence of the 
susceptibility data but is a generic fit function that should be valid 
in the close vicinity of a peak. 
We have fitted it to the peak regions of our $\chi^{\rm disc}_{\pbp}$ data. 
Each ensemble was fitted separately and we have considered several fit 
regions in temperature for each ensemble to address the systematic uncertainty
of its choice. We take the central values and the statistic errors 
from the best fit result (in terms of $\chi^2/\mathrm{dof}$) and use 
fits with $\chi^2/\mathrm{dof} < 2.5$ or with $\chi^2/\mathrm{dof} = \infty$ 
(\ie the parameters are fully constrained by the data).
Among the allowed fit results imposing the just mentioned conditions on 
the fit quality we picked the one with the maximum deviation from the 
central fit and took half of the deviation as the systematic error. 

Since we adopted a fixed scale approach the number of data points we are 
able to measure in the peak region is restricted by construction. It is even 
more restricted for the smaller pion masses since there intermediate points 
would be at large odd $N_\tau$ which are prohibitively expensive in terms 
of computing time and therefore out of reach at our present possibilities. 
Consequently, the peak region for the small pion masses is not too well 
covered by data and the fits turned out to be more difficult than those 
to the renormalized quark condensate discussed previously.

\begin{figure}[htb] 
{\centering
\hfill
\includegraphics[width=0.31\linewidth]{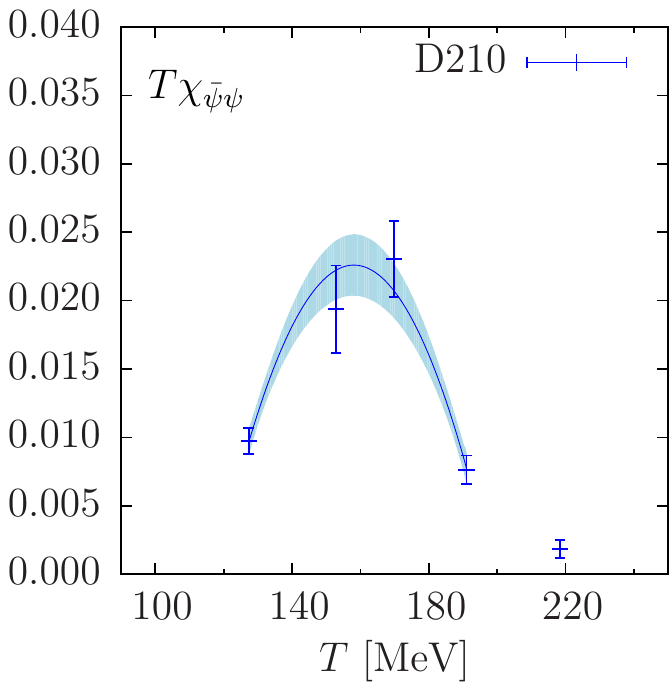}\hfill
\includegraphics[width=0.3\linewidth]{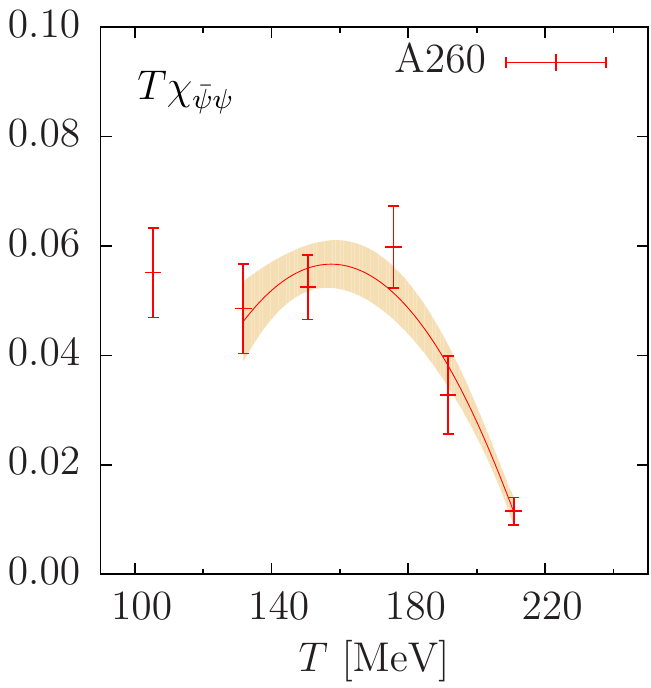}\hfill
\includegraphics[width=0.3\linewidth]{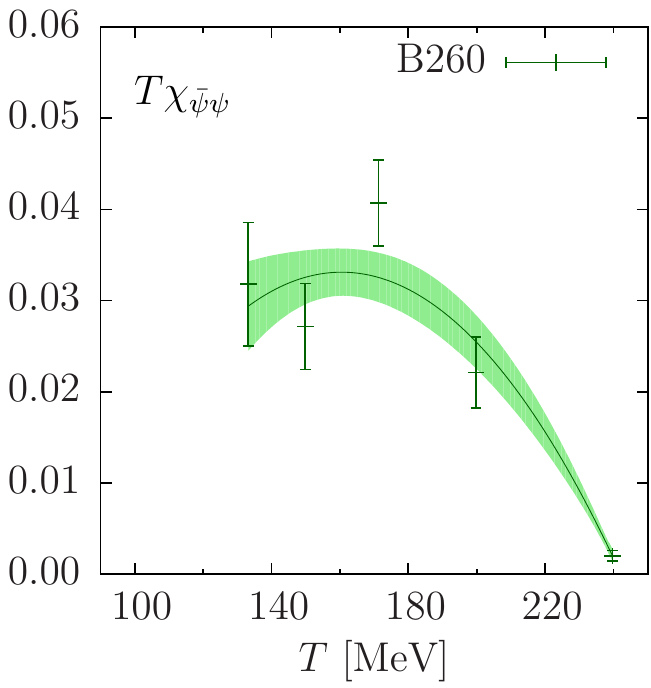}\hfill
}
\caption{Bare disconnected chiral susceptibility $T\chi^{\rm disc}_{\pbp}$ 
and fits for 
different ensembles, for the two lightest pion masses,
and different lattice spacings. Left: D210, middle: A260, right: B260 MeV. 
The best fit to the peak area for the respective ensemble is shown as a smooth 
curve together with the corresponding error envelope.}
\label{fig_suscpbp1}
\end{figure}

\begin{figure}[htb] 
{\centering
\hfill
\includegraphics[width=0.29\linewidth]{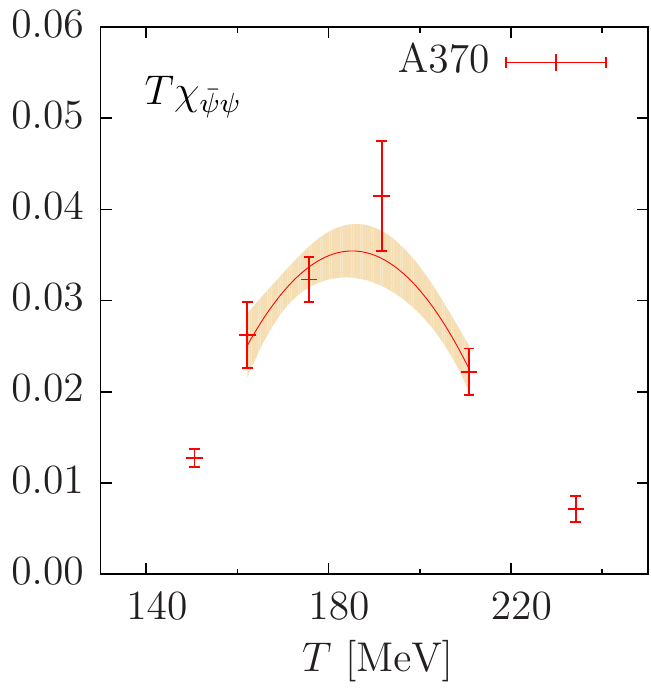}\hfill
\includegraphics[width=0.3\linewidth]{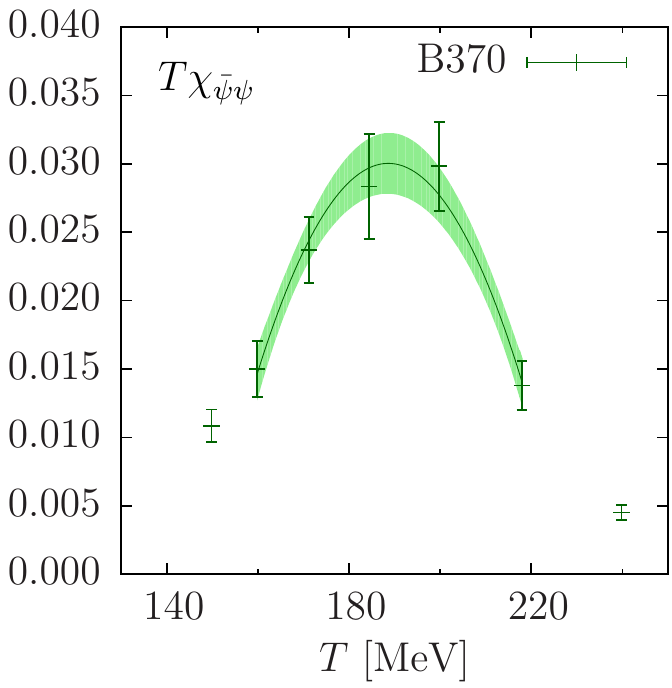}\hfill
\includegraphics[width=0.3\linewidth]{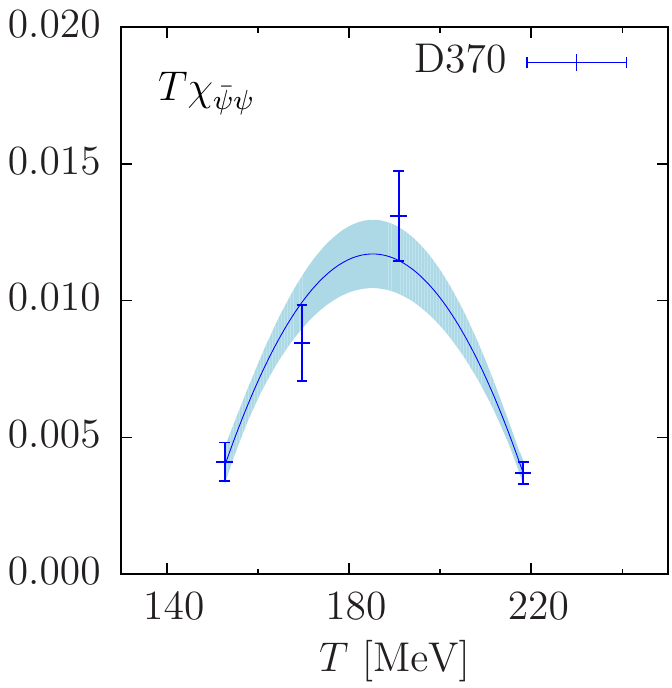}\hfill
}
\caption{Same as in Figure~\ref{fig_suscpbp1}, but for a pion mass of 370 MeV.
Left: A370, middle: B370, right: D370 MeV.}
\label{fig_suscpbp2}
\end{figure}

\begin{figure}[htb] 
{\centering
\hfill
\includegraphics[width=0.31\linewidth]{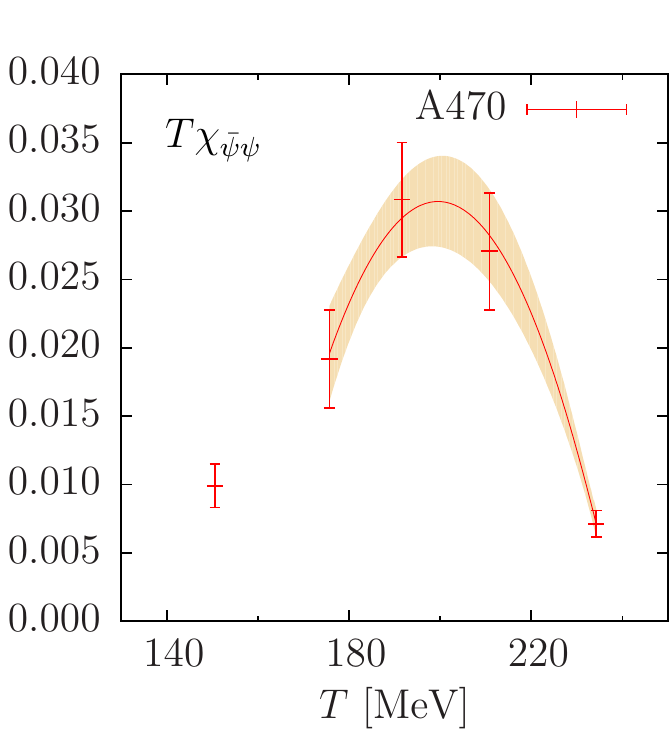}\hfill
\includegraphics[width=0.3\linewidth]{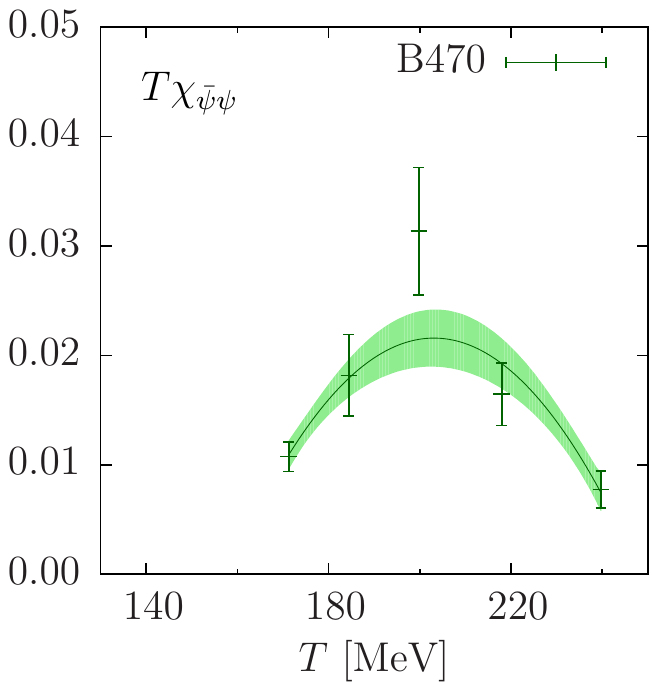}\hfill
}
\caption{Same as in Figure~\ref{fig_suscpbp1}, but for the heaviest pion mass 470~MeV.
Left: A470, right: B470 MeV.}
\label{fig_suscpbp3}
\end{figure}

In Figures~\ref{fig_suscpbp1}--\ref{fig_suscpbp3}
we show for each ensemble the best fit obtained upon varying the fit range 
in $T$.

Figure~\ref{fig_suscpbp1} shows the chiral susceptibility for the three 
ensembles with the lightest pion. While the peak is nicely visible for the 
ensemble at $\sim\!210$ MeV pion mass the maximum of the data is much more weakly 
pronounced for the two ensembles at about 260 MeV pion mass which in both cases
is mainly due to the smallest temperature points. While for the A260 ensemble 
the best fit is accordingly found discarding the smallest temperature point,
the fits for the B260 ensemble favour to include it. 

\begin{table}
\begin{center}
  \begin{tabular}{c|c|c|c|c|c|c|c}
 Ensemble & $a$ [fm] & $m_\pi$ [MeV] & $T_\chi$ [MeV] & $T_\Delta$ [MeV] &  $T^p_\Delta$ [MeV] & $T_\mathrm{deconf}$ [MeV] & $T^{\cal F}_\mathrm{deconf}$ [MeV]\\
 \hline
 D210 & 0.065 & 213 & 158(1)(4)  & 165(3)(1) & 161(2)(8) & 176(8)(8) & -  \\
 \hline
 A260 & 0.094 & 261 & 157(8)(14) & 172(2)(1) & 186(3)(8) & 188(6)(1) & 204(16)(1) \\
 B260 & 0.082 & 256 & 161(13)(2) & 177(2)(1) & 181(1)(9) & 192(9)(2) &196(13)(1) \\
 \hline
 A370 & 0.094 & 364 & 185(5)(3)  & 191(2)(0) & 202(1)(10) & 202(1)(8) &192(12)(15)  \\
 B370 & 0.082 & 372 & 189(2)(1)  & 194(2)(0) & 196(1)(8) & 201(6)(0) & 181(12)(6) \\
 D370 & 0.065 & 369 & 185(1)(3)  & 180(5)(1) & 188(2)(9) & 193(13)(2) &196(65)(20) \\
 \hline
 A470 & 0.094 & 466 & 200(4)(6)  & 193(5)(2) & 204(5)(10) & 205(4)(2)&184(15)(10)  \\
 B470 & 0.082 & 465 & 203(2)(2)  & 202(7)(1) & 204(2)(10) & 212(6)(1) &193(17)(14)  \\
\end{tabular}
\end{center}
\caption{Summary of fit-estimated pseudo-critical temperatures using fermionic 
and gluonic observables.The first error is statistical, the second systematic, see text for details }
\end{table}

\subsection{The pseudocritical line in the temperature--mass plane}

\begin{figure}[htb] 
{\centering
\includegraphics[width=10cm]{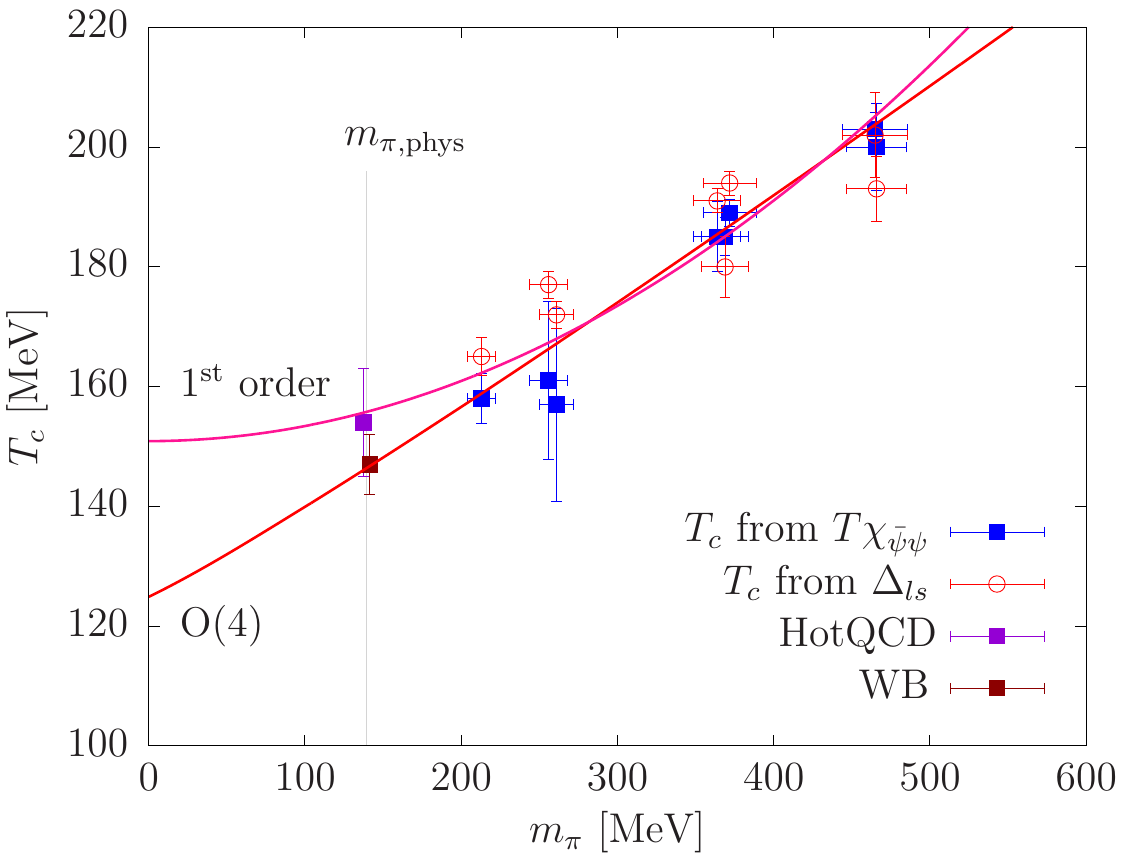}
}
\caption{ Pseudocritical temperatures as a function of the pion mass, superimposed by the curves
corresponding to first and second 
order chiral scenarios. The staggered value from Wuppertal--Budapest collaboration~\cite{BW2010}
and the coinciding result from both staggered~\cite{hotqcd2012} and domain wall~\cite{hotqcd2014} studies
of HotQCD collaboration for physical pion mass are shown as well. Data from the 
disconnected chiral susceptibility $T \chi^{\rm disc}_{\pbp}$ have been used 
in the fits. See text for discussions.}
\label{fig:chiralfit}
\end{figure}

Figure~\ref{fig:chiralfit} shows the  obtained results for $T_\chi$ and
$T_\Delta$. The estimates of $T_\chi$ and $T_\Delta$ at physical quark 
masses were obtained by the Wuppertal--Budapest collaboration~\cite{BW2010},
with staggered quark discretisations.
There is also the HotQCD result from an independent study with chiral domain wall
fermions~\cite{hotqcd2014},which is almost identical to the value from their staggered analysis~\cite{hotqcd2012}.

We fitted our values of $T_\chi$   for the two scenarios  
that have been under discussion in the context of the order of the phase transition 
in the chiral limit of the two-flavour theory~\cite{Schmidt,Cuteri}. 
First, we confirm that for these largish masses we cannot discriminate 
between different critical scenario. On the positive side, we note that either plausible
parameterizations interpolates  the data, and fares well through the results for
physical pion masses obtained with $N_f=2+1$ staggered and domain wall fermions.
We thus cross check
the results for the pseudocritical temperature provided by other groups, and,
at the same time, confirm that a dynamical charm does not
contribute to the chiral dynamics in the pseudocritical region, see Figure \ref{fig:chiralfit}. 

We close by noting two features of our results which call for further investigations.    
In the $O(4)$ scaling window we would expect 
$(T \chi^{\rm disc}_{\pbp} - T_0)/(T_\Delta - T_0) \simeq 1.5$, which is violated by our
results  once we assume $T_0 \simeq138$~MeV. 
It is an open question whether this is a  correction to $O(4)$ scaling, or if indeed the $O(4)$ scenario should be
abandoned, or maybe $T_0$ revised. 
Also, again in contrast to scaling expectation, the  two chiral temperatures  seem to get closer at large mass. This effect  
may be due to a statistical fluctuation at $m_\pi = 260$ MeV (if discard this point the trend, which anyway
small, disappears),  or, again, to large mass scaling violations. Note that 
 also the Polyakov loop pseudocritical temperature gets closer to $T_\Delta$ at larger pion masses, 
which may indeed indicate the onset of the  heavy quark regime. Obviously the validity of
the chiral scaling, and the crossover to the heavy quarks regime  are interesting questions 
but beyond the scope of the present study. We hope to be able to return to them in a future investigation.

\section{Topology and axion's properties}

Let us first briefly remind
how axion physics and topology are intimately connected. We will then
describe the results for the topological susceptibility, comment on the
scaling with the pion mass, and discuss the implications on the 
post-inflationary axion's mass bound. 

The QCD Lagrangian admits a CP violating term
\begin{equation}
\mathcal L=  {\cal{L}}_{QCD} + \theta\frac{g^2}{32 \pi^2}  F^a_{\mu \nu}\tilde {F}_a^{\mu \nu} \, ,
\end{equation}
where we recognize that 
$\dfrac{g^2}{32 \pi^2} F^a_{\mu \nu} \tilde{F}_a^{\mu \nu}$
is the topological charge density $Q(x)$.
The $\theta$ term gives an electric dipole moment to the neutron, which is
strongly constrained experimentally~\cite{Afach:2015sja} 
 leading to the bound $\theta < 10^{-10}$.
The strong CP problem consists in explaining this unnaturally small value.

An elegant solution to the strong CP problem postulates the existence of an
extra particle~\cite{PecceiQuinn,Weinberg,Wilczek},
a pseudo-Goldstone boson of the spontaneously
broken Peccei--Quinn symmetry, which couples to the QCD topological charge, with
a coupling suppressed by a scale $f_a$. thermal
 Grand Canonical partition function of QCD is now a function of $\theta$ and the temperature $T$:
\begin{equation}
\mathcal Z_{QCD} (\theta, T) = \int \!\mathcal D [\varPhi] \,e^{-T \sum_t \int d^3 x {\cal{L}}_{QCD}(\theta)} = e^{-V F(\theta, T)}
\end{equation}
and the free energy $F(\theta, T)$ is the axion potential. 

At leading order in $1/f_a$~-- well justified as $f_a \gtrsim 4 \times 10^8$~GeV~-- the axion
can be treated as an external source, and its mass is given by
\begin{equation}
m_a^2(T) f_a^2 = \frac{\partial^2 F(\theta, T)}{\partial \theta^2}\biggl|_{\theta=0} \equiv
\chi_{\rm top}(T) \; .
\end{equation}

The cumulants $C_n$  of the topological charge distribution are related 
 to the Taylor coefficients  of the expansion of the free
energy around $\theta=0$
\begin{equation}
F(\theta, T) = V \sum_{n=1}^{\infty} (-1)^{n+1} \frac{\theta^{2n}}{(2n)!} C_n,
\end{equation}
hence  higher order cumulants and their ratios 
 carry information on the axion's interactions. We will consider
higher order cumulants 
in a companion paper based on the gluonic
measurements~\cite{Trunin,inpreparation}, while in this paper we will
use our results on topological susceptibility exclusively to constrain the
(post-inflationary) axion mass.

\subsection{Topological susceptibility}

We measure the topological susceptibility following 
Refs.~\cite{KLS,Bazavov,Petreczky}. These authors start from the
continuum identity
\begin{eqnarray}
\label{eq:topqfer}
  Q(x) &=& \dfrac{g^2}{32 \pi^2} F^a_{\mu \nu} \tilde{F}_a^{\rho \sigma} \\ \nonumber
  &=& m_l \int d^4 x \bar \psi(x) \gamma_5 \psi(x)
\end{eqnarray}  
and note that this remains true on a smooth gauge 
field configuration (i.e. if lattice artifacts are small)
hence, by squaring Eq.~\eqref{eq:topqfer},
\begin{equation}
\chi_{\rm top} \equiv  \frac{\langle Q^2\rangle}{V} = m_l^2 \chi^{\rm disc}_5.
\end{equation}
It is well known that $\chi^{\rm disc}_5$ suffers from huge fluctuations. Rather than
attempting its measurement, one considers that  
when  chiral symmetry is restored,
$\chi^{\rm disc}_5 \approx \chi^{\rm disc}_{\pbp}$.

\begin{figure}
\vskip -0.5 truecm
\begin{center}
\includegraphics[width=0.31\linewidth]{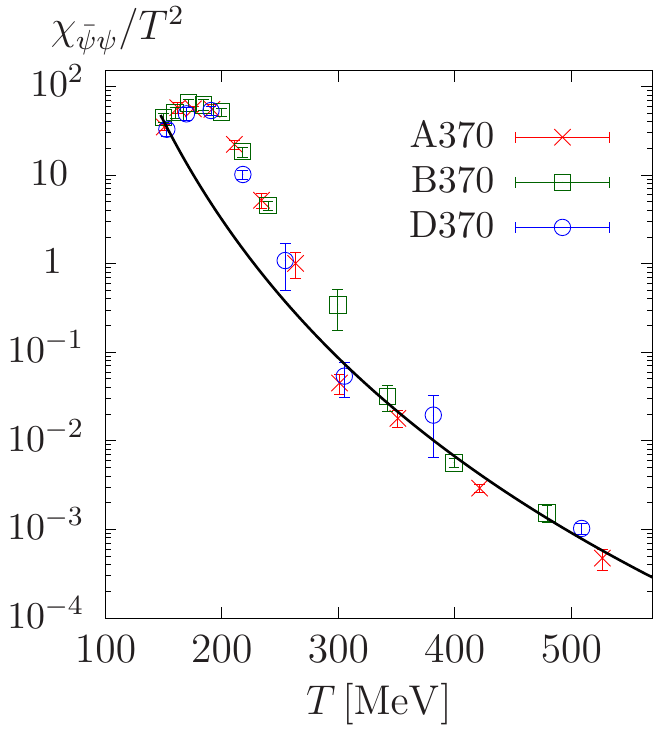}\hfill
\includegraphics[width=0.31\linewidth]{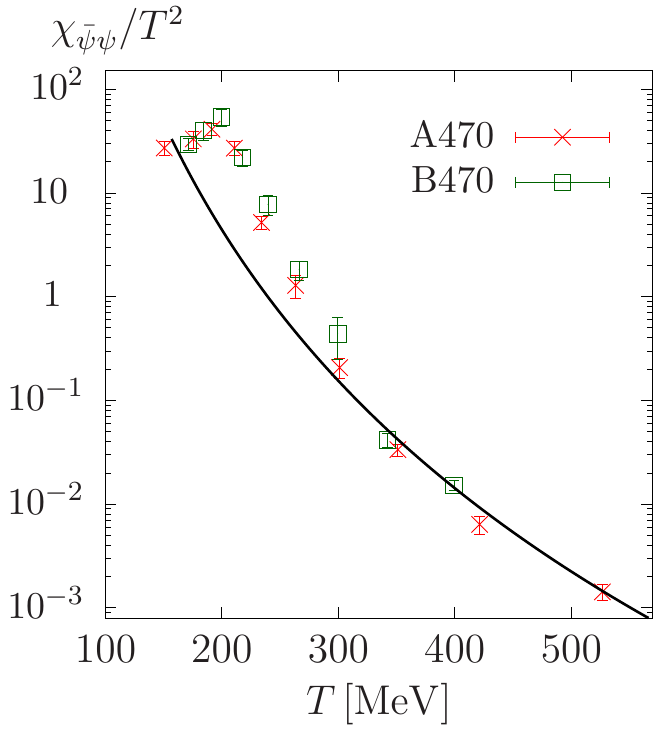}\hfill
\includegraphics[width=0.31\linewidth]{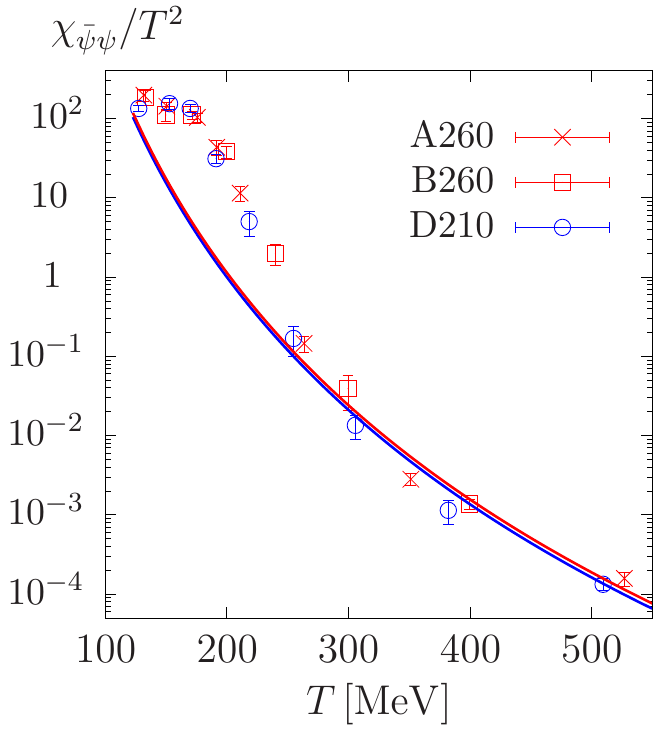}
\end{center}
\vskip -0.5 truecm
\caption{The chiral disconnected susceptibility, evaluated on the different
ensembles corresponding to  
  pion masses ranging from 210 to 470~MeV, 
and on lattices of different coarseness. For each pion mass
  the residual lattice  spacing dependence is below the statistical errors. We superimpose
the central results of the fits to the simple power-law fall off described in the text.}
\label{fig:chiallt}
\end{figure}

The main quantity of interest for our topological analysis
thus turns out to be the disconnected 
susceptibility of the chiral condensate,
$\chi^{\rm disc}_{\pbp} = \dfrac{N_\sigma^3}{T} \left( \vev{(\bar \psi \psi)^2}_l - \vev{\pbp}_l^2 \right)$
which we have already discussed within the pseudocritical region. 
For our topological study we will consider results in the extended
temperature range, 
and based on these results we evaluate the topological susceptibility
which we show in Figure~\ref{fig:chiallt}. We have grouped the results according the
mass value, and in each plot we show results for the different lattice spacing. Within 
our statistical errors we cannot see any lattice spacing dependence, and we will take
our results as continuum estimates. In the next subsection we comment
in more detail on scaling and continuum limit.

\begin{figure}
\begin{center}
\hskip -1.5truecm\includegraphics{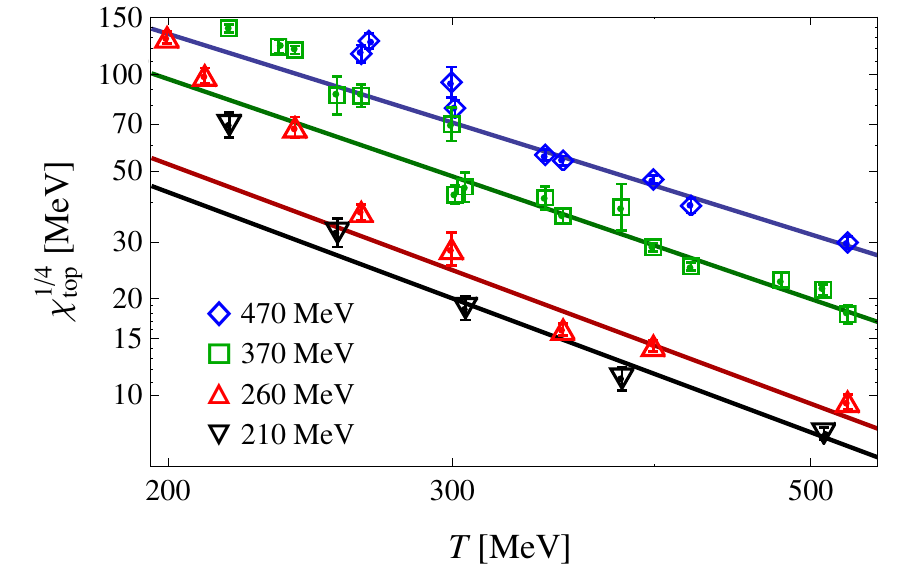}
\end{center}
\caption{The topological susceptibility from the chiral disconnected susceptibility, for different
masses at all available lattice spacings. We superimpose power-law fits $\chi_\text{top} \simeq A\, T^{-d}$ described in the text.}
\label{fig:chiallt1}
\end{figure}

In Figure~\ref{fig:chiallt1} we show all the results for the topological susceptibility
on a log-log scale in the high temperature region. 
Superimposed we show the fits to a simple
power-law  fall off  
$\chi_{\rm top}(T) = A T^{-d}$,
which describe the data for $T > 350$~MeV, with a power roughly independent on the mass.  
As already stated, we could not detect any lattice spacing dependence
within our accuracy, hence we globally fit all data belonging to the same mass for all values of lattice spacing.

\begin{figure}
\begin{center}
\includegraphics[width=0.49\linewidth]{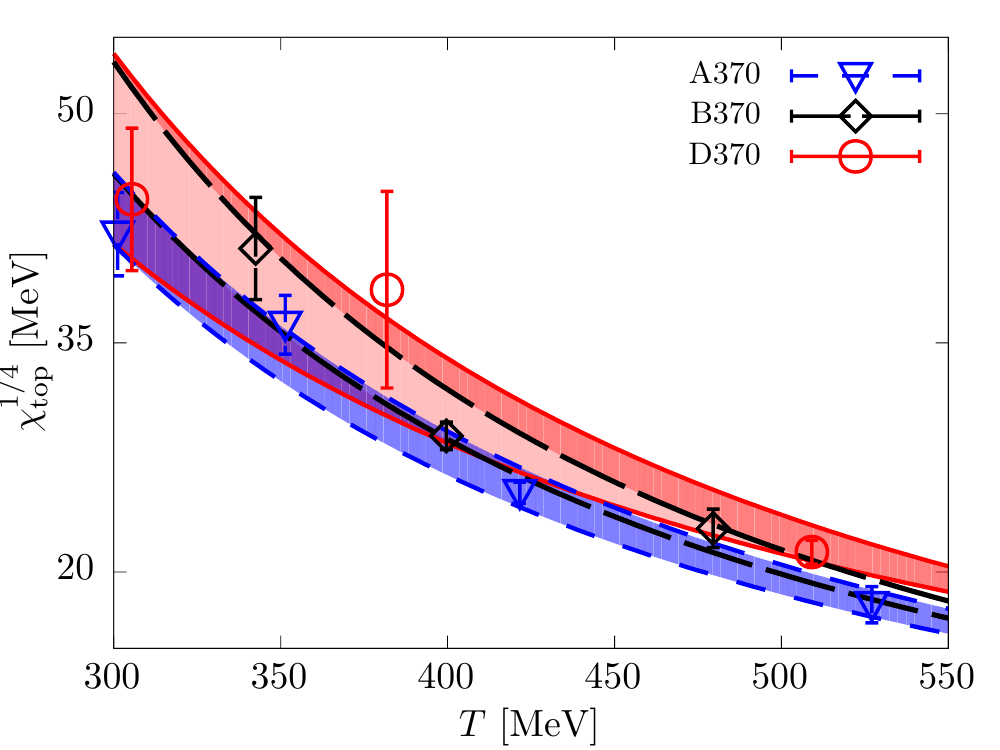}\hfill\includegraphics[width=0.49\linewidth]{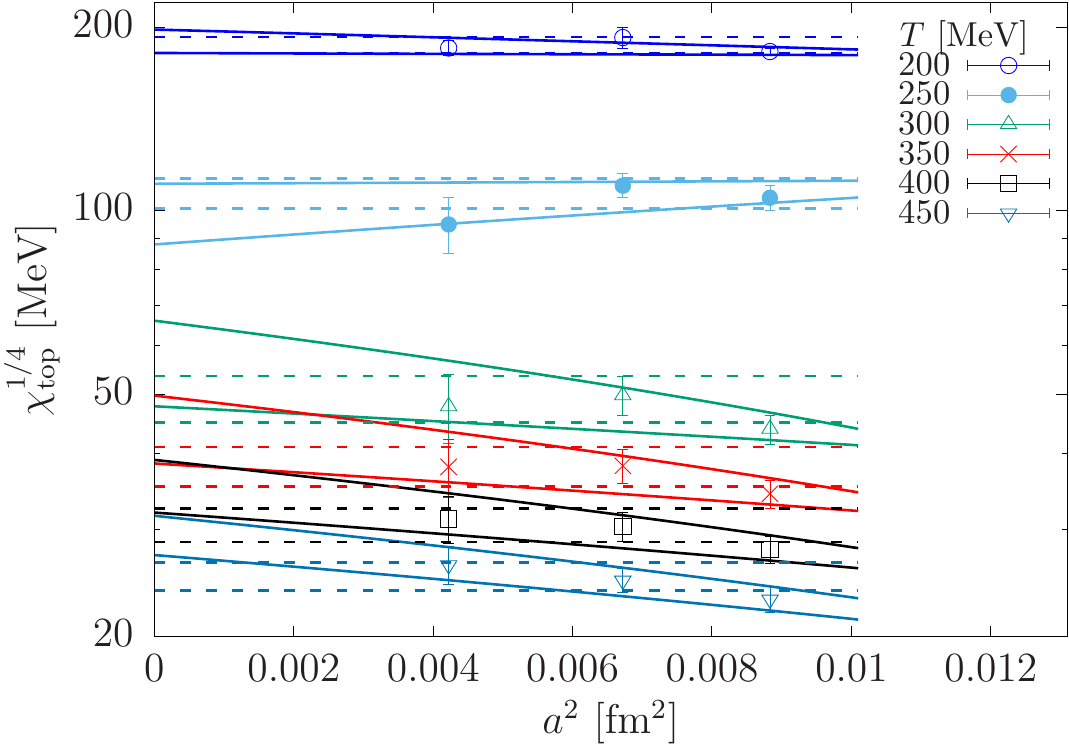}
\caption{
The topological susceptibility computed on $m_\pi=370$~MeV lattices with different spacings ({\itshape left}).
Power-law fits  for individual ensembles are superimposed, together with their errorbands.
Point-wise continuum extrapolations for interpolated $\chi_\text{top}$ ({\itshape right}). The 
interpolations needed to match temperatures for results obtained at different lattice spacings using
power-law fits (as shown in the left panel) for $T>300$ MeV and a polynomial interpolation at lower temperatures.
The dashed and solid lines correspond to zero-th and first order extrapolation in $a^2$, respectively, as described in the text.
}
\label{fig:threefits_pw}
\end{center}
\end{figure}
\begin{figure}
\includegraphics{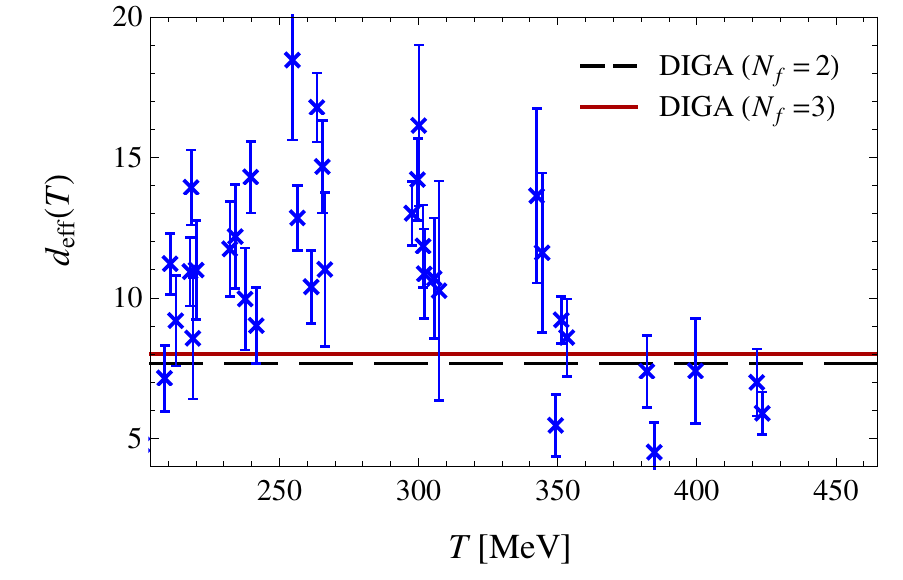}\vskip -0.5 truecm
\caption{The effective exponent describing the (local)
  power law behaviour of the topological susceptibility.
  All available pion masses and lattice spacings are used in the plot.
The results from the dilute instanton gas are shown as well.
  }
\label{fig:effexp}
\end{figure}

The fit parameters $A$ and $d$ are strongly correlated
and fit errors have little meaning~-- we merely quote the central values for $d=(6.26,6.88,7.52,7.48)$
for decreasing values of the mass $m_\pi=(470,370,260,210)$ MeV. 
To get a feeling on the error of the exponent $d$
we may explore  the functional dependence
of the topological susceptibility on the temperature
 by defining a local effective power~\cite{Aarts}
$d_\text{eff}(T) = T\,d\log \chi_{\rm top}(T)/dT$ which we show in
 Figure~\ref{fig:effexp}. In doing so, we do not distinguish different masses
 and spacings: as the log derivative
 is rather noisy, it is not possible to detect any trend with
 mass and spacing and we simply plot all the results together.
The average result is broadly consistent with  the fits,
and apparently approaches a constant value  above $T \sim 350$~MeV, while
 the spread of the exponents could be taken as an estimate of the
 error on the power $d$ of the fall-off. 

Concerning the apparent constant asymptotic behaviour, we expect that
at very large temperatures  the partition function is
described by a dilute gas of instantons and anti-instantons (DIGA),
which leads to a well defined prediction for the fall off of the
topological susceptibility, also shown in Figure~\ref{fig:effexp}. 
Apparently the DIGA result is approached already for these temperatures:
however,  it is not clear
from the plot whether such  decreasing trend with temperature has reached its
nearly asymptotic value (modulo small corrections),
or, rather, whether the apparent coincidence with the DIGA value is
accidental, and restricted to a limited range of temperatures. Only simulations
for larger temperatures  can settle this issue.

Further, we note that the DIGA result is approached from above: the exponent 
has a larger value for smaller
temperature~-- see again
Figure~\ref{fig:chiallt1}.
Very interestingly, it has recently been proposed~\cite{Larsen} that close to $T_c$
the dilute instanton gas changes to an ensemble of instantons and dyons,
and the signature for this should be a faster decrease of the topological
susceptibility close to $T_c$.

\subsection{Details on lattice artifacts}

We focus on
the results for $m_\pi = 370$ MeV, where we have results at the three lattice
spacing. Firstly, we performed individual power law fits
that we are going to use as interpolators for $T > 300$ MeV. We display
the fits for
$m_l^2 \chi^{\rm disc}_{\pbp}  \simeq \chi_{\rm top}$  in the left diagram of Figure~\ref{fig:threefits_pw}.
Next, we show in Figure~\ref{fig:threefits_pw}, right, the lattice spacing
dependence of    $m_l^2 \chi^{\rm disc}_{\pbp}$ for different temperatures, mostly above
 $T_c$ (remember that
this quantity is a proxy for  $\chi_{\rm top}$ only in the symmetric phase). 
The required interpolations to obtain matching temperatures have used a polynomial
interpolation around and slightly above
the pseudocritical temperature, and the power law fits just described for $T > 300$ MeV. 
We then consider two continuum extrapolations, 
$m_l^2 \chi^{\rm disc}_{\pbp}(a^2) = m_l^2 \chi^{\rm disc}_{\pbp} + O(a^2)$ 
and 
$m_l^2 \chi^{\rm disc}_{\pbp}(a^2) = m_l^2 \chi^{\rm disc}_{\pbp} + K a^ 2 + O(a^4)$ , 
i.e.
we truncate  the $a^2$ series for 
$m_l^2 \chi^{\rm disc}_{\pbp}$ at the zero-th (dashed lines) or at a first order (solid lines in Figure~\ref{fig:threefits_pw}).
In the first case we consider the two points at the smallest lattice spacings,
and we note that ensemble A is anyway well described within errors,
in the second case we
use the three points we have at our disposal. All fits fare nicely through the data, with
continuum extrapolated results for $m_l^2 \chi^{\rm disc}_{\pbp}$ in good agreement between each other, 
and with the results
at the smallest lattice spacing. A word of warning is however in order: it  is obvious that a 
robust  continuum extrapolation would require at least another lattice spacing. At the moment, within
the given restrictions and this caveat issued, we conclude that we observe good scaling
and that using results on finite lattice as estimates of continuum ones is legitimate. 
We also note that the scaling pattern we
observe is quite comparable with those reported by the ETMC at zero temperature 
in their study of the topological susceptibility from the twisted mass Dirac operator spectrum~\cite{Cichy:2013rra}.

\subsection{Pion mass dependence and rescaling}
Our results have been obtained with pion masses ranging from $210$ till 
$470$ MeV, above the physical pion mass. The same dilute instanton
gas model has a prediction for the mass scaling which may be used
to extrapolate our results to the physical pion mass: $\chi_{\rm top} \propto m_\pi^4$. 
We note that this leading mass dependence is more general than DIGA 
and  simply follows from the analyticity
of the chiral condensate in the chiral limit above $T_c$. In fact, 
taking $\langle\bar \psi \psi\rangle = \sum_{n=0} a_n m_l^{2n+1}$ in the symmetric phase,
the total susceptibility
is an even series in the quark mass 
\begin{equation}
\chi = \frac{V}{T} \frac{\partial}{\partial m_l} \vev \pbp 
\equiv \chi^{\rm disc}_{\pbp} + \chi^{\rm conn}_{\pbp} = \sum_{n=0}  a_n m_l^{2n}.
\end{equation}
Barring unexpected cancellation we may assume that the same holds for the connected
and disconnected susceptibilities separately, hence
\begin{equation}
\chi_{\rm top} =  m_l^2 \chi^{\rm disc}_{\pbp} = \sum_{n=0} a_n m_\pi^{4(n+1)}.
\end{equation}
An exact DIGA form would imply that the leading order is exact,
i.e. that the disconnected chiral
susceptibility does not depend on the pion mass in the mass range considered. 
Our results do show a mass dependence, not surprisingly, given the largish masses
which we are using~-- smaller masses are probably needed to get rid of the
subleading mass corrections.  In this first study, rather than attempting
an extrapolation in mass  we content ourselves with the simple rescaling 
dictated by the leading order, as done first in~\cite{Bonati}.

\begin{figure}
\includegraphics{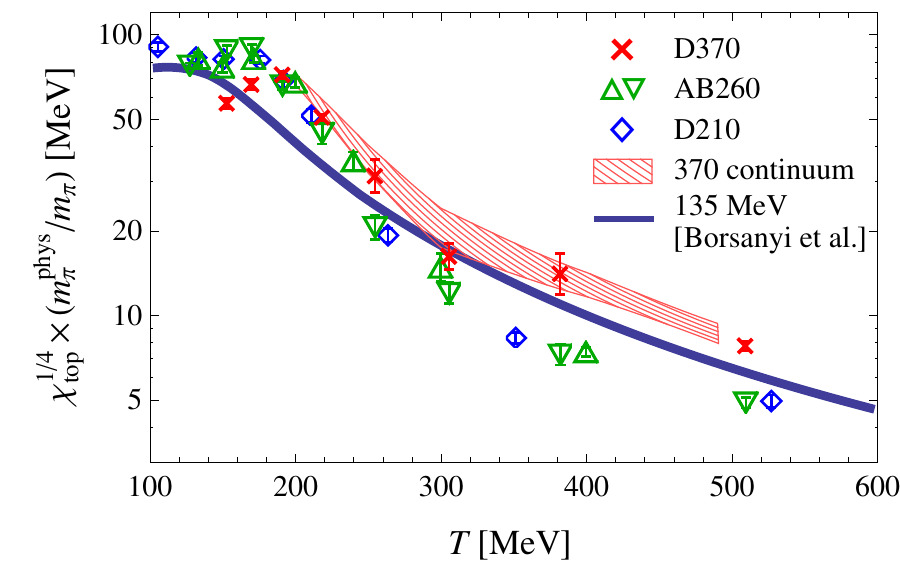}
 \vskip -0.5 truecm
\caption{The fourth root of topological susceptibility versus the temperature for 
three different pion masses, rescaled to the physical pion mass according to $\chi_{\rm top} \propto m_\pi^4$.
For $m_\pi=370$~MeV continuum band the solid line extrapolations from the right panel of Figure~\ref{fig:threefits_pw} are taken.
We also superimpose the tabulated results (Table S7) from~\cite{Borsanyi}.} 
\label{fig:topsusc}
\end{figure}

In Figure~\ref{fig:topsusc}  we present the results for 
the topological susceptibility obtained by rescaling the data for different pion masses to the physical pion mass
according to the leading scaling prescription.
For $m_\pi=370$~MeV, the continuum result from the first-order extrapolation in $a^2$ is also given, which consists with the D-ensemble datapoints within errorbars.
In the same diagram we reproduce
the results from  Table S7 of Ref.~\cite{Borsanyi} obtained with physical
quark masses for a comparison. As we have already discussed, the trend close to
$T_c$ is different, while  there is a broad agreement
at high temperatures -- we will discuss below the implications of the
residual differences on the axion mass. 
A similar agreement holds with the results of Ref.~\cite{Petreczky}.

\subsection{Axions}

\begin{figure}[htb] 
\includegraphics{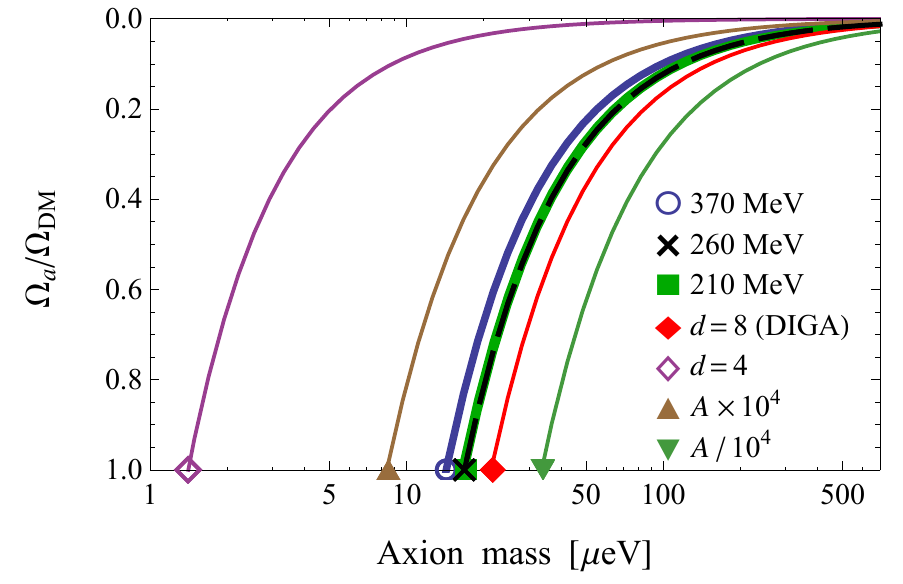}
\caption{The axion contribution to Dark Matter versus the axion mass: the first three lines
show our results for three pion masses 370, 260, and 210 MeV, respectively,  all rescaled to the physical pion mass according to $\chi_{\rm top} \propto m_\pi^4$.
The other lines are mock data meant to study the sensitivity of the 210~MeV curve to
the fit parameters of the topological susceptibility $\chi_\text{top} \simeq A\, T^{-d}$.}
\label{Fig:axions-fraction}
\end{figure}

A power-law decay for the topological susceptibility as noted by many authors  opens a very
interesting possibility: a safe extrapolation to very large temperatures. 
This feature has been exploited in  applications to axion physics, briefly 
reminded in the Introduction. In a nutshell (see e.g Refs.~\cite{turner}), when the axion mass is of the order of the inverse of the Hubble parameter, the axion starts to oscillate:  
$3 H(T) = m_a(T) = \sqrt{\chi_\text{top}(T)}/f_a$. 
$f_a$ can be traded for the zero temperature topological susceptibility 
and the axion mass via
$\chi_\text{top} = f_a^2 m_a^2$, hence the knowledge of the temperature dependence  of the
topological susceptibility suffices to determine the time  of the beginning of oscillation. At this time,
the energy density $\rho_a(T)$ of the oscillating axion field is the same as a collection of axions
at rest $\rho_a(T) \approx 1/2 m_a^2(T) f_a^2 \theta^2$, and the number 
density  $n_a (T) = \rho_a(T)/m_a(T)$ can be estimated
as $n_a(T) \approx 1/2 m_a(T) f_a^2 \theta^2$ .
The axion-to-entropy ratio remains constant after the beginning of the oscillations,
so the present mass density of axions is $\rho_{a,0} = \dfrac{n_a}{s} m_a s_0$,
where $s,s_0$ are the entropies at time $T$,
and of today, and $\Omega_a = \dfrac {\rho_{a,0}}{\rho_c}$,
with $\rho_c$ the critical density.  
To obtain simple expressions in closed form we employ the power-law parameterization
$g^*(T) = 50.8 \,(T/(\text{MeV}))^{0.053}$
for the number of relativistic degrees of freedom $g^*(T)$ 
entering the Hubble parameter and entropy density, which
reproduces the results up to a few percent in the temperature interval $800< T <1500$~MeV.
Following e.g.~\cite{turner}  we finally arrive
at $\rho_a(m_a) \propto m_a^{-\frac{3.053 + d/2}{2.027 + d/2}}$, where $d > 0$
defines the power law decay of the topological susceptibility at high temperatures discussed
above, $\chi_\text{top} \simeq A\, T^{-d}$, and we use the latest PDG results  for the required 
astrophysical constants~\cite{PDG}.
We present the results in a graphic form in Figure~\ref{Fig:axions-fraction}: we plot 
the axion's fractional contribution to Dark Matter 
versus the axion mass for various situations. Similar analyses have been presented
in previous works, and we confirm to a large extent their conclusions~\cite{Borsanyi,Petreczky,moore}.
The first three lines are obtained from our results, rescaled to the physical
pion mass. We have included errors for the lowest mass, which are not visible in the graph.
The discrepancy between the results for the two lowest pion masses, and for a pion of $370$~MeV
may be ascribed  to violations to the mass scaling discussed above, and are anyway
small at practical level. Clearly, as we know, further uncertainties might be hidden 
in the results~-- to estimate their impact we plot a few mock curves 
on the basis of 210~MeV data: first, we keep the 
same amplitudes as the one we have measured, but choose the DIGA exponent. Second, we choose
a small exponent similar to the one reported in~\cite{Bonati}. To study the effect of the amplitude,
we fix the exponent to the one we have measured at 210~MeV, and consider the fourth root of the amplitude ten times larger or
smaller.  In all cases  the intercept with the abscissa axis (overclosure bound) defines the absolute lower
bound for the axion mass: for our results, this gives $m_a \simeq 20 \mu$eV. 
 This limit becomes more stringent if we assume that the axions only
contribute to a fraction of DM~-- the results can be read off the plot from the intercept of
the desired fraction and the corresponding curves. 
Clearly the bounds are robust against 'small' changes of parameters. However a significant 
variability  remains. In particular, as also noted in Ref.~\cite{Villadoro},
slower decays (as those observed by us using the gluonic definition of the topological susceptibility) would considerably lower the axion bound.

\section{Discussion}

We have studied chiral and topological properties of $N_f = 2+1+1$
QCD with twisted mass Wilson fermions. The strange and charm
mass have their physical values, while for the light pion  we
have considered four different masses.

We have identified the pseudocritical temperature as a function
of the pion mass. We have found that an extrapolation to the
pseudocritical  temperature for 
physical pion is robust with respect to different assumption for
the universality class of the two flavor massless theory,
and agrees well with previous estimates. This serves as a sanity
check of for the twisted mass chiral dynamics around $T_c$ 
and in addition it confirms
the irrelevance of a dynamical charm in the transition region. 

We have measured the topological susceptibility in the range 
150$-$500  MeV and found a rather fast decrease 
with temperature in the range  $T_c < T \lesssim 350$~MeV. 
This feature is predicted in recent instanton-dyon models~\cite{Larsen},
however it has not been observed
in other recent lattice studies. Such fast decrease may also been 
understood within the framework of the QCD magnetic Equation of 
State~\cite{Ejiri}, as it is known that around $T_c$ the 
disconnected susceptibility almost saturates the total susceptibility 
$\partial \langle\bar \psi \psi\rangle / \partial m$. 
Since the total susceptibility in the symmetric phase behaves
as $1/(T - T_c)^\gamma$ with $\gamma \simeq 1$  (with 
a weak mass dependence due to Griffith analyticity) it is rather 
natural to expect a fast decrease of the topological susceptibility 
leading to an apparently large exponent in a simple power-law 
parameterization $\/T^\alpha$.
This scenario should be checked by directly computing the regular contribution
to the susceptibility, and most importantly the $\gamma_5$ susceptibility,
which would allow the computation of the topological susceptibility
without relying on the restoration of the chiral symmetry.
At higher temperatures the contribution from the regular part becomes
significant and the results approach the DIGA behaviour from above~-- although,
as we have already mentioned it is hard to exclude a continual decreasing trend which would
bring the results well below that.

This work might be extended along several directions: firstly~-- and obviously~--
the results should be further pushed towards lower masses. The first zero temperature
results  by the ETM collaboration with $N_f = 2 +1 + 1$ and  a physical 
pion mass  have appeared recently~\cite{etmc-phys},
and we hope to be able to extend them to high temperature.  Second, we would
like to clarify in detail the source of discrepancies among different methods.
Although there are several reasons to believe that these discrepancies may
be traced back to lattice artifacts, we are still in need of a clear evidence.
In a companion paper we will present results with a gluonic method obtained
on the same lattices~\cite{inpreparation}, and a study with the overlap operator is in progress~\cite{inpreparation}.
Last but not the least, we hope to go beyond the second order cumulant studied here in order to learn more on the axion potential. 

\vskip 0.1 truecm
We are grateful to the Supercomputing Center of Lomonosov Moscow State
University, to the HybriLIT group of JINR, to the HLRN 
and to CINECA (INFN-CINECA 
agreement) for computational resources. E.-M.~I. and M.~P.~L. wish to 
thank the Director and the Staff at the European Center for Theoretical 
Nuclear Physics, Trento, for their kind support and hospitality, and the 
Theory Group at Justus Liebig Universit\"at in Giessen. M.~P.~L. thanks 
the Bogoliubov Laboratory for Theoretical Physics of JINR Dubna 
and the Galileo Galilei Institute for Theoretical Physics for 
hospitality. This work is partially supported by the COST action THOR and RFBR grant 18-02-01107. A.~T. acknowledges support from the BASIS foundation.

\end{document}